\documentclass[aps,prb,twocolumn,amsmath,amssymb,superscriptaddress,showpacs]{revtex4-1}
\usepackage{graphicx}
\usepackage{dcolumn}
\usepackage[usenames]{color}
\usepackage{bm}
\usepackage[normalem]{ulem}

\newcommand{\bti}{Bi$_2$Te$_2$I$_2$}
\newcommand{\btb}{Bi$_2$Te$_2$Br$_2$}
\newcommand{\btc}{Bi$_2$Te$_2$Cl$_2$}
\newcommand{\btx}{Bi$_2$Te$_2X_2$}

\newcommand{\bise}{Bi$_2$Se$_3$}
\newcommand{\bite}{Bi$_2$Te$_3$}

\begin{document}

\title{2D and 3D topological phases in BiTe$X$ compounds}

\author{S. V. Eremeev}
 \affiliation{Tomsk State University, 634050 Tomsk, Russia}
 \affiliation{Saint Petersburg State University, 198504 Saint
Petersburg,  Russia} \affiliation{Institute of Strength Physics and
Materials Science, 634055 Tomsk, Russia}
 \affiliation{Donostia
International Physics Center, 20018 San Sebasti\'{a}n/Donostia,
Spain}

\author{I. A. Nechaev}
 \affiliation{Centro de F\'{i}sica de Materiales CFM - MPC and Centro
Mixto CSIC-UPV/EHU, 20018 San Sebasti\'{a}n/Donostia, Spain}
 \affiliation{Tomsk State University, 634050 Tomsk, Russia}
 \affiliation{Saint Petersburg State University, 198504 Saint
Petersburg,  Russia}

\author{E. V. Chulkov}
 \affiliation{Centro de F\'{i}sica de Materiales CFM - MPC and Centro
Mixto CSIC-UPV/EHU, 20018 San Sebasti\'{a}n/Donostia, Spain}
 \affiliation{Tomsk State University, 634050 Tomsk, Russia}
 \affiliation{Saint Petersburg State University, 198504 Saint
Petersburg,  Russia} \affiliation{Donostia International Physics
Center, 20018 San Sebasti\'{a}n/Donostia, Spain}
 \affiliation{Departamento de F\'{i}sica de Materiales UPV/EHU,
Facultad de Ciencias Qu\'{i}micas, UPV/EHU, Apdo. 1072, 20080 San
Sebasti\'{a}n/Donostia, Spain}

\date{\today}

\begin{abstract}
Recently, it was shown that quantum spin Hall insulator (QSHI) phase
with a gap wide enough for practical applications can be realized in
the ultra thin films constructed from two inversely stacked
structural elements of trivial band insulator BiTeI. Here, we study
the edge states in the free-standing \bti\ sextuple layer (SL) and
the electronic structure of the \bti\ SL on the natural BiTeI
substrate. We show that the topological properties of the \bti\ SL
on this substrate keep $\mathbb Z_2$ invariant. We also demonstrate
that ultra thin centrosymmetric films constructed in the similar
manner but from related material BiTeBr are trivial band insulators
up to five-SL film thickness. In contrast to \bti\ for which the
stacking of nontrivial SLs in 3D limit gives a strong topological
insulator (TI) phase, strong TI is realized in 3D \btb\ in spite of
the SL is trivial. For the last material of the BiTe$X$
($X$=I,Br,Cl) series, BiTeCl, both 2D and 3D centrosymmetric phases
are characterized by topologically trivial band structure.
\end{abstract}

\pacs{71.15.-m, 71.18.+y, 73.22.-f}

\maketitle

\section{Introduction}

Starting from the theoretical predictions by Kane and Mele
\cite{KaneMele_PRL2005} and Bernevig et al.
\cite{Bernevig_PRL2006,Bernevig_Sci2006}, the $\mathbb Z_2$
two-dimensional topological insulator (2D TI) or the QSHI phase, in
which spin-helical gapless edge states counter-propagate along the
boundary with opposite spins, providing quantum spin Hall effect
(QSHE)\cite{KaneMele_PRL2005} with quantized spin-Hall conductance,
attract considerable attention of researchers. After experimental
observations of the QSHE in HgTe/CdTe and InAs/GaSb quantum
wells,\cite{Konig,Knez} a number of 2D TIs were theoretically
proposed.\cite{Ren_rev} These proposals were mainly based on the
layers of bismuth, graphene, or heavy elements analogs of graphene,
implying the gap tuning by strain, adatoms deposition, chemical
functionalization, growing on substrates, or sandwiching. In
contrast to 3D TIs, for which the existence of topologically
protected Dirac surface states was found in a wide variety of
materials, the edge states in 2D TIs by now were directly observed
experimentally in a limited number of systems like Bi(111)-bilayer
islands on a Bi-crystal surface \cite{Drozdov}, in Bi(110)\cite{Lu}
and Sb(111)\cite{Kim-Yeom} thin films, and step edges in ZrTe$_5$,
\cite{Li_ZrTe5,Wu_ZrTe5} while the QSHE was confirmed in the quantum
wells with tiny band gaps only,\cite{Ren_rev} where a topologically
protected state does not survive at temperatures above 10 K. Thus,
the goal that remains to be actual so far is to search for robust
and easily fabricated new 2D TIs with a sufficiently large band gap
providing edge states accessible to experimental probes at room
temperature.


Recently, it was suggested a distinct way to design novel
topological systems on the base of two-dimensional materials
consisting of layered band insulators.\cite{Nechaev} It was
demonstrated that these systems can be realized under normal
conditions in thin films comprising van-der-Waals coupled structure
elements (trilayers, TLs) of the giant-Rashba semiconductor BiTeI,
which belongs to the intensively investigated chalcogenides BiTe$X$
($X$=Cl, Br, and I).
\cite{Ishizaka, Eremeev_PRL2012, Crepaldi,Landolt_PRL2012, Eremeev_JETPLett2012, Eremeev_NJP2013,Landolt_NJP,Sakano_PRL2013, Rusinov_PRB2013, Mauchain, Butler, TournierColletta, Tran, Fiedler, Fiedler_PRB2015, Rusinov_NJP2016, Maass_NatComm_2016}

BiTe$X$ compounds have a hexagonal non-centrosymmetric crystal
structure built of ionically-bonded $X$-Bi-Te TL stacked along the
hexagonal $z$ axis. In addition to the large bulk and surface Rashba
splitting in these materials, a single BiTe$X$ trilayer holds the
giant Rashba-split states as well.\cite{Eremeev_SRep_TI-BTX} Among
BiTe$X$ materials, BiTeI demonstrate the largest bulk Rashba
splitting, while BiTeCl stands out for its isotropic spin-split
metallic surface state lying deep inside the bulk band gap, which is
the biggest one in the BiTe$X$ series. Generally, the BiTe$X$
surfaces, which can be Te- or $X$-terminated, can possess,
respectively, electron- or hole-like Rashba-split surface states
that emerge by splitting off from the lowest conduction (highest
valence) band, owing to the negative (positive) surface potential
bending.\cite{Eremeev_PRL2012} However, it turned out that the
surface of BiTeI obtained by natural cleavage of single crystals
grown by the Bridgman method always hold both types of the surfaces
states due to two different type domains (Te- and I-terminated),
inhomogeneously distributed over the surface.
\cite{Crepaldi,Landolt_PRL2012,Butler,TournierColletta,Fiedler,Fiedler_PRB2015,Maass_NatComm_2016}
These surface domains are the consequence of a large number of
randomly distributed bulk stacking faults, in which the TLs stacking
order is inverted.\cite{Fiedler,Fiedler_PRB2015,Maass_NatComm_2016}
The calculated bulk stacking fault formation energy in the BiTeI
bulk, 1~meV, is much smaller than for BiTeBr (46 meV) and BiTeCl (60
meV)\cite{Fiedler_PRB2015} that explains the existence of the mixed
domain termination on the BiTeI surface in contrast to the BiTeBr
and BiTeCl (0001) surfaces characterized by single polar domains.

It was demonstrated \cite{Nechaev} that a centrosymmetric sextuple
layer (SL) constructed from two BiTeI TLs with facing Te-layer sides
is a 2D TI with the inverted gap of about 60 meV at $\bar{\Gamma}$,
which is sufficiently enough for room temperature spintronics
applications. The gap inversion occurs due to the
bonding-antibonding splitting between one of the Te-related valence
bands (VBs) and one of the conduction bands (CBs) formed by Bi
orbitals. The films of a several SL thickness demonstrate rapid
decrease in the gap width with the number of SLs along with
oscillating behavior in $\mathbb{Z}_2$  topological invariant like
in thin films of the \bite\ family (see, e.g.,
Refs.~\onlinecite{Bihlmayer_Book_TI, Foerster_PRB_2015,
Foerster_PRB_2016, NeKras_PRBR_2016}). The corresponding bulk system
composed of SLs turned out to be a strong 3D TI (hereafter referred
to as \bti). It is energetically unfavourable by only 0.5 meV
compared with the non-centrosymmetric BiTeI. It should be noted that
due to the stacking faults mentioned above crystals of BiTeI grown
by the Bridgman method already contain the desired SLs, and that the
inverted stacking can be experimentally observed and controllably
manufactured.

Most of earlier theoretically predicted 2D TIs were considered as
free-standing systems. However, for practical applications, a
freestanding 2D TI must be placed or grown on appropriate substrate
which, in general, will influence the intrinsic topological
properties of thin films due to interfacial and proximity effects
\cite{Menshov} such as mismatch in the lattice constants and charge
transfer at the interface of diverse materials.
\cite{Takayama,Hirahara,Chuang_NanoLett2014} Therefore, suitable
substrates to support room-temperature TIs are important for
potential device applications. Several 2D TI systems, mostly
Bi-based layers on silicon (or SiC) substrate were theoretically
suggested.\cite{Zhou_PNAS,Zhou_SciRep,Hsu,Huang_PRB2013,Huang_PRB2014,Song,Chuang_PRB2016}
However, they have not been experimentally realized up to now. The
exception is the honeycomb Bi on
Au/Si(111)-($\sqrt3$$\times$$\sqrt3$) substrate structure, which was
successfully grown as confirmed by LEED and STM
measurements.\cite{Chuang_PRB2016} Nevertheless, the QSHE has not
been proved to be observable in this system so far. In this regard,
the \bti-SL 2D TI has a significant advantage, since it has a
natural substrate -- the non-centrosymmetric BiTeI band insulator,
which guarantees an absence of both lattice mismatch and interface
charge transfer effects, since there is no work function difference
between the substrate and the 2D TI adlayer.

In this paper, using first-principles electronic structure
calculations, we examine the edge states of the \bti\ 2D TI of
one-SL thickness for different edge orientations. We also consider
the effect of the BiTeI substrate on the electronic structure of
this 2D TI. Finally, we study the topological properties of the
related 2D and 3D centrosymmetric phases \btb\ and \btc.

\begin{figure}[t]
  \includegraphics[width=\columnwidth]{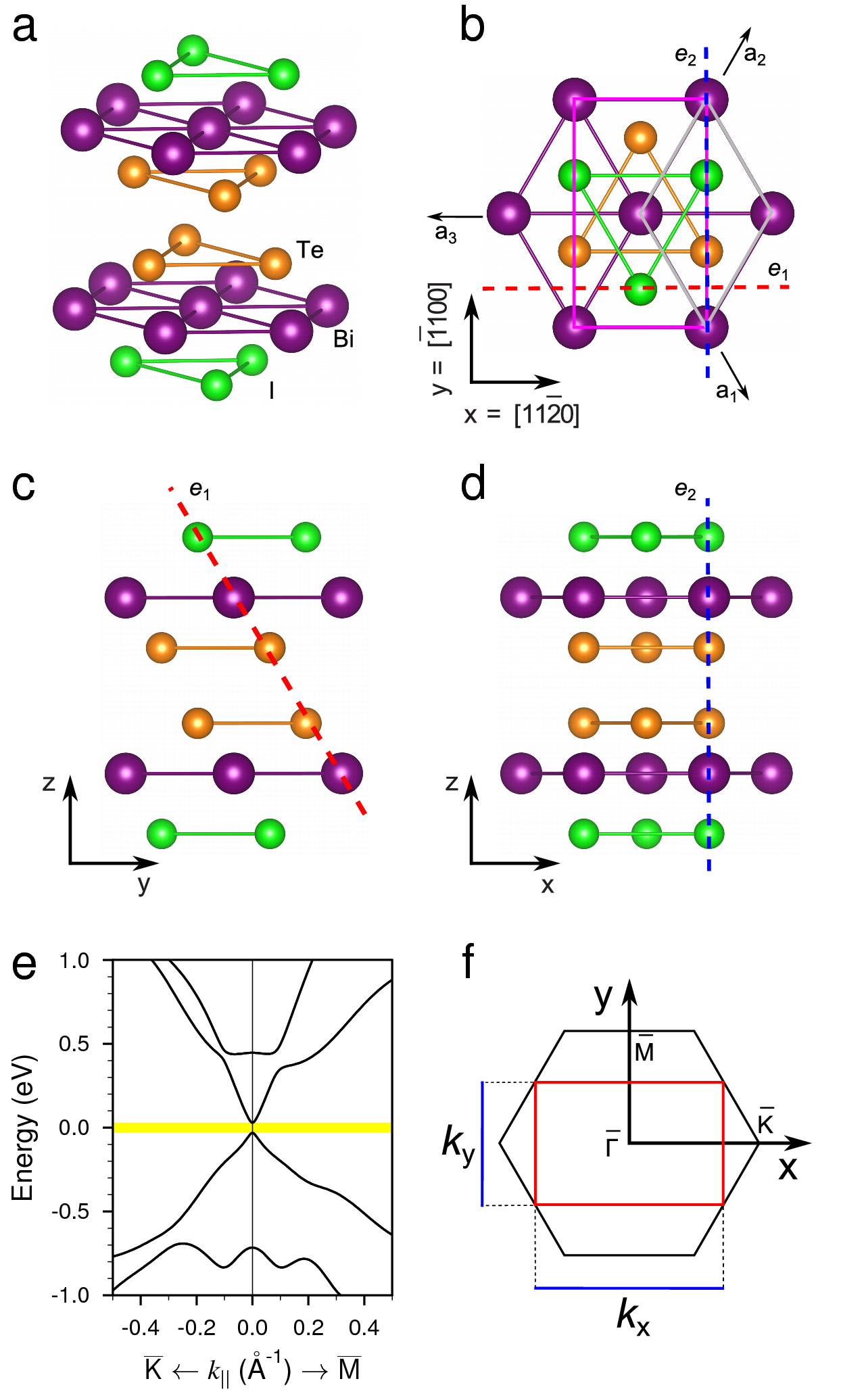}
\caption{Atomic structure of the \bti\ one-SL film (a) with its top
view (b) and side views in $y-z$ and $x-z$ planes (c and d,
respectively). $e_1$ and $e_2$ dashed lines (red and blue,
respectively) show cleavage planes for considered stoichiometric
edges. Light-gray rhombus and magenta rectangle in (b) mark the unit
cell and related rectangular cell by repeating of which in $x$ or
$y$ direction we construct different nanoribbons. (e) Band structure
of the free-standing \bti\ SL along $\bar\Gamma-\bar{\rm K}$ and
$\bar\Gamma-\bar{\rm M}$ directions of the 2D hexagonal Brillouin
zone (BZ); yellow stripe shows the region of the band inverted gap.
Black hexagon in (f) shows 2D hexagonal BZ; red rectangle shows 2D
BZ for rectangular cell and blue lines indicate its projections onto
1D BZs along $k_x$ and $k_y$ axes. }
 \label{struct}
\end{figure}

\begin{figure*}[t]
\includegraphics[width=\textwidth]{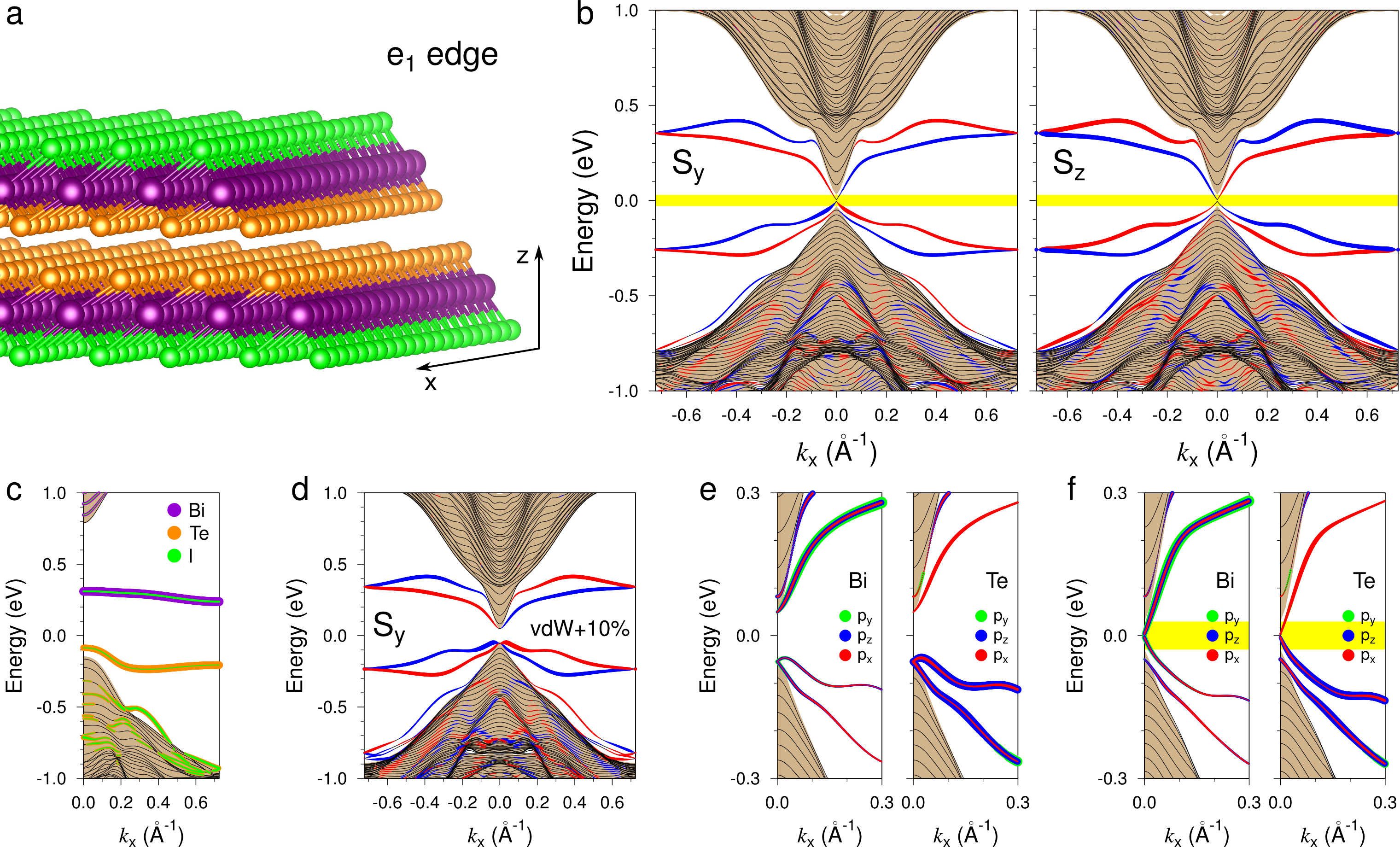}
\caption{(a) Atomic structure of the $e_1$ edge. (b) Electronic
spectrum and spin polarization for localized states at the edge;
red/blue circles represent positive/negative sign of the spin
component for localized states. (c) Atom-projected localization of
the trivial dangling bond states as calculated with the switched off
SOC. (d) Electronic structure of the $e_1$ edge of the topologically
trivial \bti\ film with vdW spacing expanded by 10\%. Orbital
character of the localized states for Bi (left panel) and Te (right
panel) contributions for the trivial (vdW+10\%) system (e) and the
\bti\ QSH insulator (f). }
 \label{edge1}
\end{figure*}

\begin{figure*}[t]
   \includegraphics[width=\textwidth]{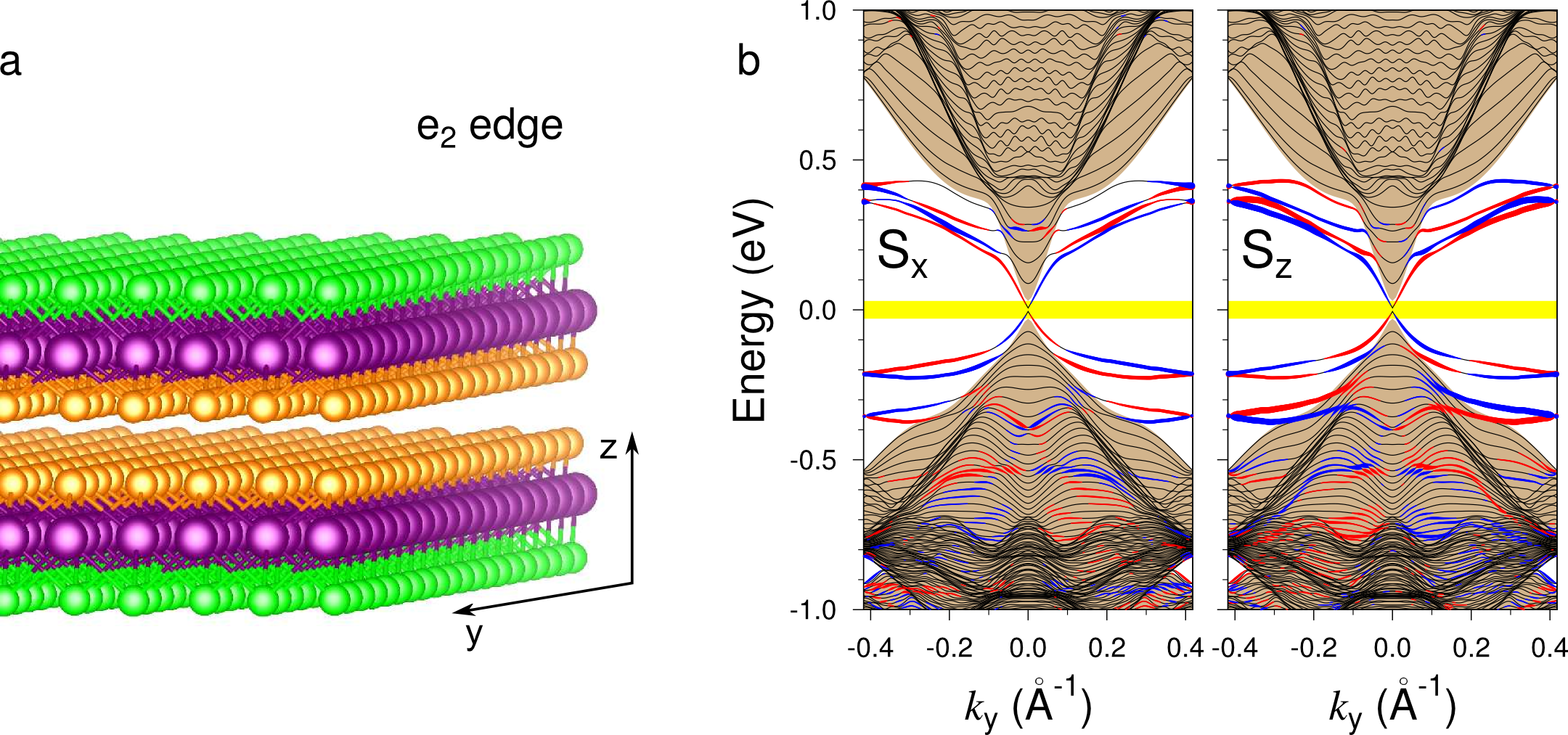}
\caption{Atomic structure (a) and electronic spectrum with spin
polarization (b) for the $e_2$ edge of \bti\ nanoribbon.}
 \label{edge2}
\end{figure*}

\section{Methods}

Calculations were carried out within generalized gradient
approximation (GGA) with the projector augmented-wave method
\cite{PAW1,PAW2} as realized in the Vienna Ab Initio Simulation
Package (VASP).\cite{VASP1,VASP2} DFT-D3 van der Walls (vdW)
correction \cite{Grimme} was applied for accurate structure
optimization. Bulk lattice parameters and atomic positions of
Bi$_2$Te$_2X_2$ phases (the 164 (P$\bar{3}$m1) space group) were
optimized. The optimized $a$ and $c$ parameters for \bti\ bulk have
been obtained to equal 4.354 and 13.421 \AA, respectively, and they
were found to be 4.474 and 11.676 \AA\ and 4.288 and 12.155 \AA\ for
\btb\ and \btc, respectively. For all \btx\ bulk phases, the $a$
parameter is slightly larger than that in respective
non-centrosymmetric BiTe$X$ compounds. The Bi-Te and Bi-$X$
interlayer spacings in \btx\ bulk are also close to those in
BiTe$X$; Te-Te($X$-$X$) interlayer distances in \btx\ bulk phases
vary in the range of $X$=I, Br, Cl as 2.655(3.094), 2.456(2.521),
and 2.796(2.623) \AA. For single free-standing \btx\ SLs, we
performed additional structural optimization. A small contraction
was obtained in the Te-Te distance, and a small expansion was found
in the outer Bi-$X$ spacing, while Te-Bi interlayer spacing remained
unchanged as compared to the respective bulk values. The $\mathbb
Z_2$ topological invariant was calculated by using the method based
on tracking the evolution of hybrid Wannier functions realized in
Z2Pack \cite{Soluyanov}.

\section{Results and discussion}

\subsection{Edge states in \bti-SL nanoribbons}

First we studied the energies of edges with different geometry to
find out the most stable edge for the \bti-SL nanoribbon. We have
considered stoichiometric mutually perpendicular cleavage planes
$e_1$ and $e_2$, see Fig~\ref{struct}(b--d). The $e_1$ plane is
parallel to the $[\bar1100]$ ($x$) axis and crosses (0001) ($xy$)
plane at an angle of 56.96$^\circ$. The $e_2$ plane is parallel to
the $[11\bar20]$ ($y$) axis and passes perpendicular to the (0001)
plane. As expected from the effective continuous model,\cite{Linder}
the decay depth of the topological edge states in 2D TI should be
much larger than that for the Dirac surface states in 3D TI.
Consequently, to avoid the finite size effect, the nanoribbon width
should be chosen as large as possible. We have constructed
nanoribbons bounded by the two parallel $e_1$ and $e_2$ cleavage
planes of the width of $\sim$150 \AA. We found that the $e_1$ edge
is by $\sim$170 meV/\AA$^2$ more favorable than the $e_2$ one. The
stability of the $e_1$ edge looks reasonable, because the preferred
cleavage plane is the one, which minimizes the number of broken
bonds. The square of the $e_1$ plane is about 1.5 times smaller than
that of the $e_2$ plane, while the number of broken bonds at the
$e_2$ plane is two times bigger as compared to $e_1$ (8 and 4,
respectively), and thus the density of the broken bonds (per unit
square) is smaller for the $e_1$ edge. Note that the stoichiometric
cleavage plane, similar to $e_1$ was found to be stable for the
Bi$_2$Te(Se)$_3$ single quintuple layer (QL).\cite{Virk}

Next, we examine the electronic structure of the ribbons with the
$e_1$ and $e_2$ edges. Fig.~\ref{struct}(e) shows the spectrum of
the \bti\ film of single SL thickness calculated along high-symmetry
directions of the hexagonal Brillouin zone [Fig.~\ref{struct}(f)].
As was demonstrated in Ref.~\onlinecite{Nechaev} this film has an
inverted gap of $\sim$ 60 meV at the Fermi level, and in this gap
(yellow stripe) we expect to find one-dimensional topological
spin-helical states localized at the edge of the nanoribbon. When
the $e_1$ ($e_2$) edge is formed, the 2D BZ of the \bti-SL is
projected onto 1D BZ along the $k_x$ ($k_y$) direction
[Fig.~\ref{struct}(f)].

The calculated electronic structure of the $e_1$ ribbon
(Fig.~\ref{edge1}(a)) is shown in Fig.~\ref{edge1}(b), where left
and right panels demonstrate the same electronic spectrum but
different projections of the spin expectation value for the
localized states, $S_y$ and $S_z$, which are orthogonal to the wave
vector $k_x$ ($S_x$ component of the spin is zero). They are
calculated as a sum over atomic contributions from five near-edge
atomic layers. As can be seen in Fig.~\ref{edge1}(b), the spectrum
has a gap at the Dirac point (of few meV) as the result of
hybridization between edge states localized at the opposite edges of
the ribbon despite we have chosen quite wide ribbon. Another
observation is that beside the $\bar\Gamma$ Dirac state the band
structures also show the presence of a number of spin-polarized
resonant states mainly in the valence band region as well as two
trivial Rashba-split dangling bond states in the gap region far from
the $\bar\Gamma$ point. These states are originated from the broken
bonds at the edge. With switched off spin-orbit coupling (SOC) these
dangling bond states appear as two almost dispersionless bands
[Fig.~\ref{edge1}(c)]. The occupied and unoccupied bands have an
only small iodine contribution and the former band is mainly
contributed by Te orbitals, while the latter is primarily formed by
the Bi orbitals. Both bands have $p_{yz}$ character over 1D BZ that
reflects the fact the dangling bonds are oriented perpendicular to
the $e_1$ cleavage plane.

As was shown in Ref.~\onlinecite{Nechaev}, the vdW interaction
between TLs that form the \bti-SL is crucial to realize QSHI phase.
The gap in the SL spectrum closes at increasing the vdW spacing by
5\% and further increase in the vdW spacing reopens trivial gap. At
10\% expansion, the spectrum has the trivial gap of the width
comparable to that in the equilibrium band-inverted
SL.\cite{Nechaev} Constructing the nanoribbon with the vdW spacing
expanded by 10\%, we found merely two dangling bond states acquired
the Rashba splitting (Fig.~\ref{edge1}(d), $S_y$ spin component is
only shown). At large $k_x$ ($>$0.3 \AA$^{-1}$) despite the spin
splitting both bands keep their $p_{yz}$ character, whereas at small
$k_x$ [Fig.~\ref{edge1}(e)] the orbital character is changed due to
emergence of $p_x$ orbitals and the appearance of Bi $p_{yx}$ states
in the occupied band along with Te $p_x$ states in the unoccupied
band. Comparing the orbital character of the Rashba-split dangling
bond states near the $\bar\Gamma$ point at the edge in trivial
(vdW-expanded) SL with that in the QSHI [Fig.~\ref{edge1}(f)], one
can conclude that the emergent Dirac states have the atom-type
contributions and the orbital compositions similar to that of
trivial states, with which the Dirac states hybridize strongly.

\begin{figure*}[t]
 \includegraphics[width=\textwidth]{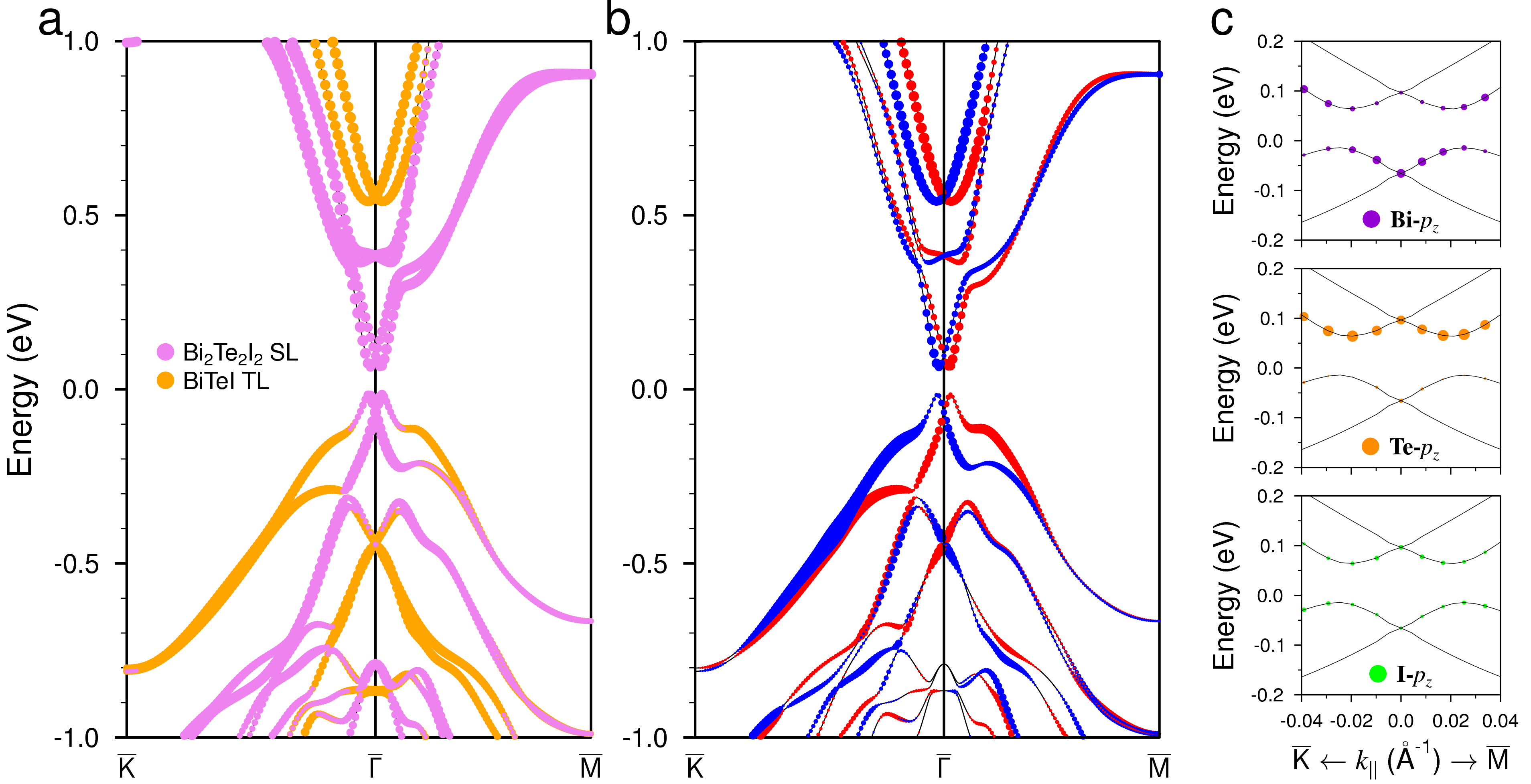}
\caption{Layer- (a) and spin-resolved (b) spectrum for
\bti-SL/BiTeI-TL film. (c) Orbital-resolved band structure in the
vicinity of the $\bar\Gamma$-gap.}
  \label{1TL}
\end{figure*}

\begin{figure*}[t]
  \includegraphics[width=\textwidth]{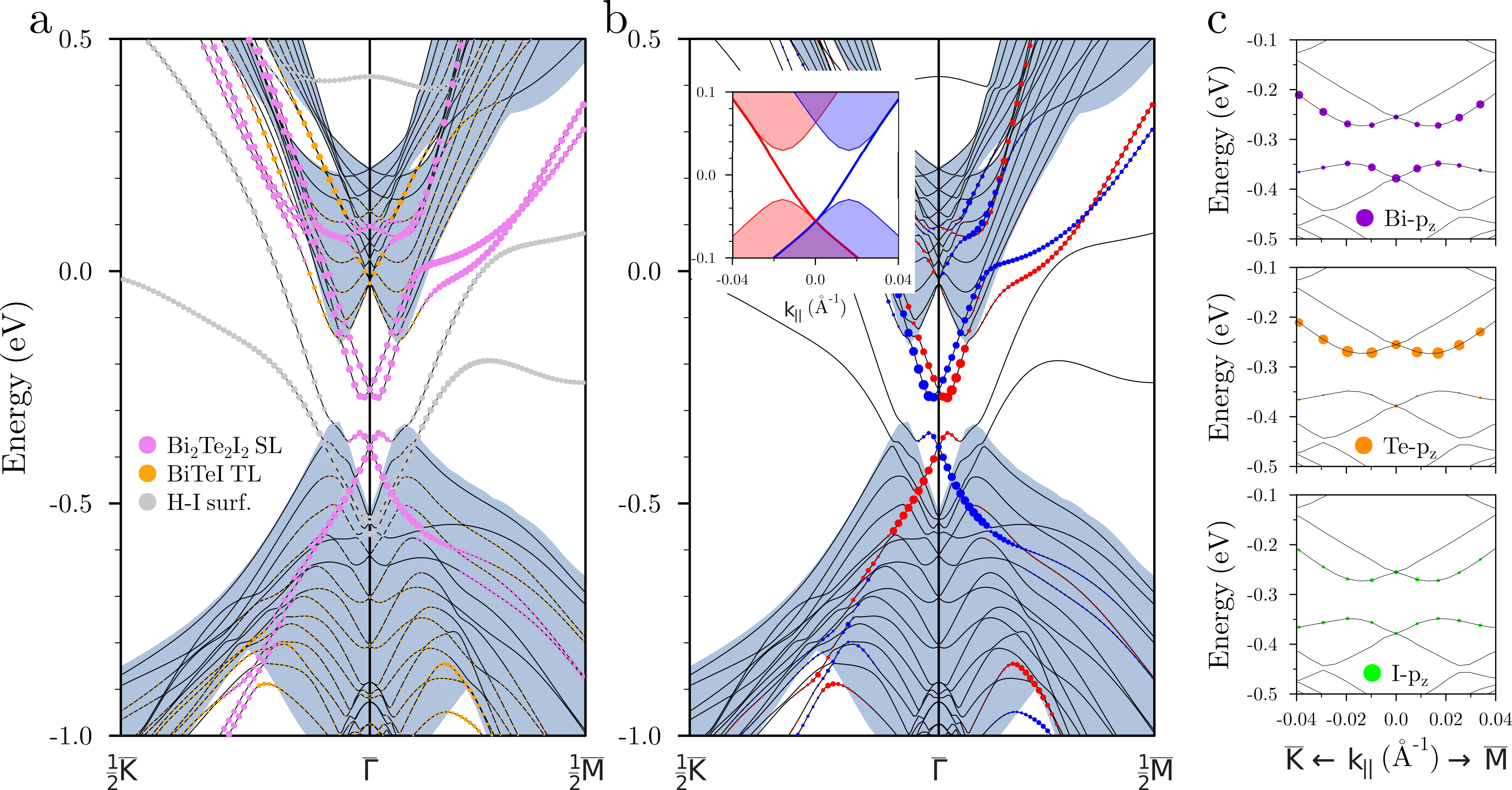}
\caption{Layer- (a) and spin-resolved (b) spectrum for \bti-SL@BiTeI
structure. (c) Orbital-resolved band structure in the vicinity of
the $\bar\Gamma$-gap.}
  \label{8TL}
\end{figure*}

In the ribbon with the less stable $e_2$ edge [Fig.~\ref{edge2}(a)],
the spectrum and different spin projections, $S_x$ and $S_z$ ($S_y$
is zero in this case) of which are presented in Fig.~\ref{edge2}(b),
additional expected trivial Rashba-split dangling bond states are
emerged in the projected gap. The number of such bands is two times
larger as compared to the spectrum of the $e_1$ edge owing to the
fact that twice as many bonds break at the formation of this edge.
However, as in the nanoribbon with $e_1$ edge, the dangling bond
states lie far from the $\bar\Gamma$ gap region (yellow stripe)
where topological Dirac state resides.

\subsection{\bti-SL 2D TI on BiTeI band insulator substrate}

As we pointed above, one of the main challenge for 2D TIs is
appropriate substrate, and the advantage of \bti\ is that it has a
natural substrate -- BiTeI band insulator, which guarantees the
absence of both the lattice mismatch and the interface charge
transfer effects. Since the charge transfer is mostly a local
effect, we consider first the structure composed of \bti-SL and a
single BiTeI TL as [I-Bi-Te--Te-Bi-I]$_{\rm SL}$-[Te-Bi-I]$_{\rm
TL}$. The calculated band spectrum of this structure is shown in
Fig.~\ref{1TL}. As can be seen in Fig.~\ref{1TL}(a), the states
forming a gap in this system belong to the \bti-SL, while the states
localized in the BiTeI-TL lie far from the gap edges, especially in
the unoccupied part of the spectrum. Since the
[I-Bi-Te--Te-Bi-I]$_{\rm SL}$-[Te-Bi-I]$_{\rm TL}$ system lacks the
inversion symmetry, the states possess the Rashba spin-splitting
(Fig.~\ref{1TL}(b)). The splitting is almost isotropic with respect
to $k_{||}$, and splitting parameters are $\Delta k$=0.026(0.020)
\AA$^{-1}$ and $\Delta E$=0.048(0.029) eV for valence (conduction)
band. Beside the gap in \bti\ states acquires the Rashba splitting,
its width becomes larger. In the free-standing inversion symmetric
\bti\, the gap is of 60 meV (Fig.~\ref{struct}(e)), and in
[I-Bi-Te--Te-Bi-I]$_{\rm SL}$-[Te-Bi-I]$_{\rm TL}$ the absolute gap
$E_g$ equals 78 meV (the $\bar\Gamma$-gap is $E_g$+$\Delta E^{\rm
VB}$+$\Delta E^{\rm CB}$). However, despite the Rashba splitting the
gap-edge states remain inverted, as in the inversion symmetric
\bti.\cite{Nechaev} In the vicinity of $\bar\Gamma$, the Bi
$p_z$-orbitals mainly localized in the valence band, while Te $p_z$
orbitals dominate in the lowest conduction band (Fig.~\ref{1TL}(c)).
The calculation of $\mathbb Z_2$ for [I-Bi-Te--Te-Bi-I]$_{\rm
SL}$-[Te-Bi-I]$_{\rm TL}$ structure resulted in non-trivial
topological invariant.

Next we increased the number of TLs in the slab so that the BiTeI
part of the structure was sufficient to reproduce the BiTeI bulk. As
in earlier works
\cite{Eremeev_PRL2012,Landolt_PRL2012,Eremeev_JETPLett2012} we used
BiTeI slab of 8 TLs thickness with the back, iodine-terminated, surface
passivated by hydrogen. As can be seen in Fig.~\ref{8TL}(a,b), in
\bti-SL@BiTeI heterostructure the \bti\ states lie within the BiTeI
bulk gap. The gap in \bti\ states $E_g$=74 meV is almost the same as
in case of single TL BiTeI substrate but spin-splitting parameters
became slightly smaller: $\Delta k$=0.018(0.014) \AA$^{-1}$ and
$\Delta E$=0.030(0.018) eV for valence (conduction) band. Such
small alterations in the \bti\ spectrum in the vicinity of
$\bar\Gamma$ do not affect the gap inversion as can be seen in
Fig.~\ref{8TL}(c). As before, the gap edges are mainly formed by
$p_z$ orbitals of Bi (VB) and Te (CB) atoms.

Thus, 2D TI \bti\ survives on BiTeI substrate. This is due to the
fact that \bti\ and BiTeI are actually the same materials, which
differ in TLs staking order only. It guarantees the absence of the
interface potential because the interface [I-Bi-Te--Te-Bi-I]$_{\rm
SL}$-[Te-Bi-I]-[Te-Bi-I]-... can be regarded as the structure
[I-Bi-Te]-[Te-Bi-I]-[Te-Bi-I]-[Te-Bi-I]-..., i.e., as the BiTeI
surface, where uppermost TL has inverted atomic layer sequence. The
splitting in the spin subbands in the 2D TI \bti\ spectrum on the
BiTeI substrate will result in the opposite $k_\|$-shifting of the
corresponding spin branches of the Dirac state in the 1D spectrum as
schematically shown in the inset in Fig.~\ref{8TL}(b). Wherein, the
Dirac point that is in the middle of the gap in the case of
free-standing SL might be shifted downward to the region of the
two-dimensional states.

\subsection{Search for 2D and 3D topological phases in \btb\ and \btc}

Finally, we consider 2D and 3D centrosymmetric phases constructed
from TLs of the related band insulators BiTeBr and BiTeCl. Despite the
bulk stacking faults, which is a key element for centrosymmetric
BiTe$X$ pases, are missing in BiTeBr and BiTeCl obtained by the
Bridgeman method as follows from the
experiment\cite{Fiedler_PRB2015} demonstrating the absence of the
mixed-terminated surfaces in these materials, however, one can
assume that centrosymmetric films of BiTeBr and BiTeCl can, in
principle, be epitaxially grown.

Calculating the total energies of free-standing SLs of \btb\ and
\btc\, we found that they have lower (by 98 and 91 meV,
respectively) energies than BiTeBr and BiTeCl films of double TL
thickness have. However, the bulk phases composed of \btb\ and \btc\
SLs are less favorable as compared to corresponding BiTeBr and
BiTeCl non-centrosymmetric bulk crystals. The energy losing for
\btb\ with respect to BiTeBr is 20.75 meV per BiTeI fu and it is
26.35 meV/fu for \btc\ system in comparison with BiTeCl.

\begin{figure}
  \includegraphics[width=\columnwidth]{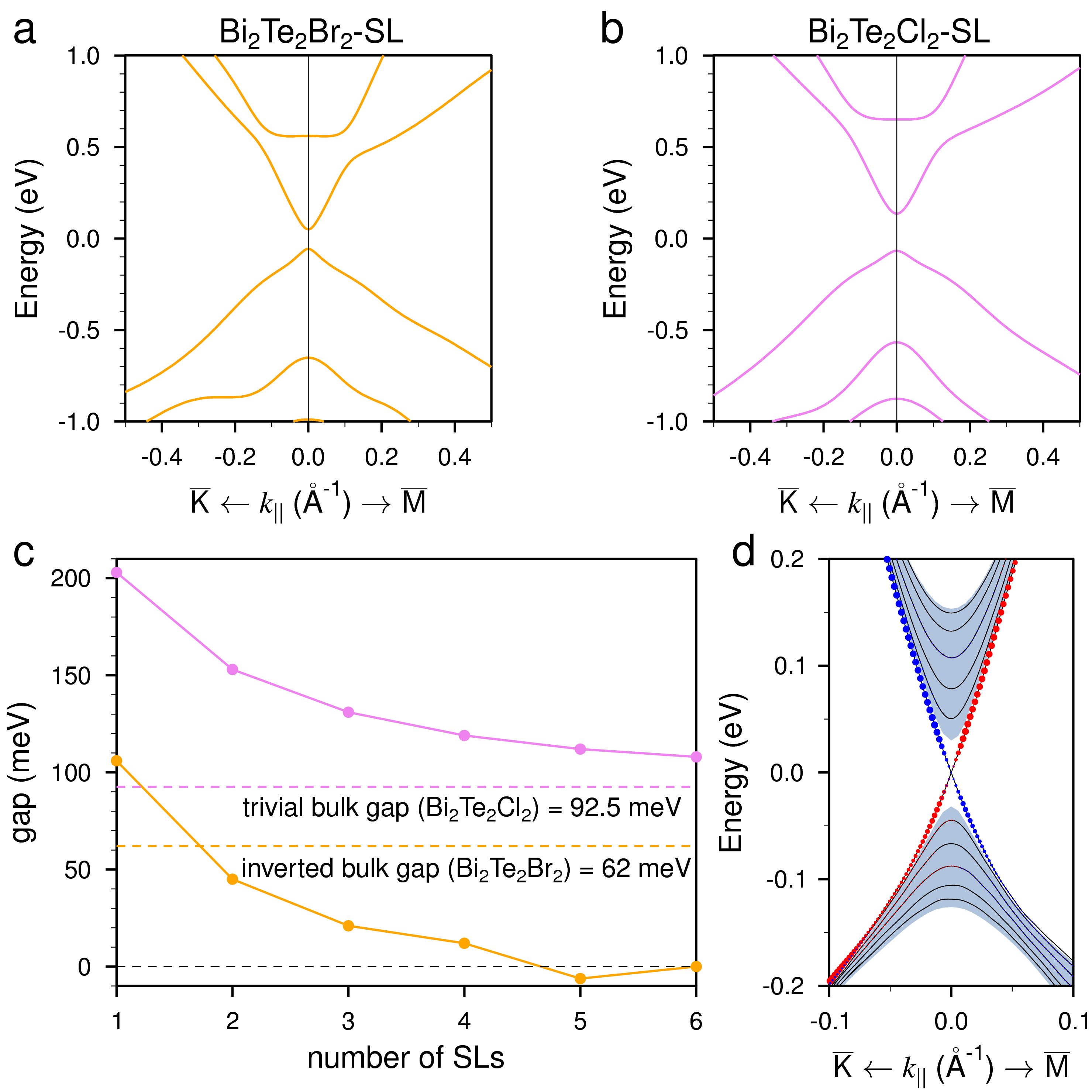}
\caption{Band spectra of free-standing \btb\ (a) and \btc\ (b) SLs.
The dependence of the $\bar\Gamma$-gap width on the number of SLs in
\btb\ and \btc\ thin films (c). The sign of the gap is positive for
topologically trivial film and negative for non-trivial one. (d)
Surface electronic structure of the 3D TI \btb.}
  \label{BTB-BTC}
\end{figure}

The band spectra of \btb\ and \btc\ single, free-standing SLs
[Fig.~\ref{BTB-BTC}(a,b)], in general, are similar to that of \bti,
however, they have larger band gaps, 106 and 203 meV, respectively,
and, according to $\mathbb Z_2$ calculations, these centrosymmetric
\btb\ and \btc\ SLs are topologically trivial.

Next, like the case of \bti, we addressed the electronic structure
of the \btb\ and \btc\ bulk. We revealed that \btb\ has an inverted
band gap of 62 meV while \btc\ bulk is a trivial band insulator with
the gap of 92.5 meV and thus it is trivial in both 2D and 3D phases.

Considering the dependence of the $\bar\Gamma$-gap and the
topological invariant on the film thickness for \btb\ and \btc\
[Fig.~\ref{BTB-BTC}(c)], we found that in contrast to \bti\, where
along with a sharp decrease in the value of the gap, the oscillating
behavior of $\mathbb Z_2$ with the number of SLs was obtained,
\cite{Nechaev} the \btc\ demonstrates a smooth decrease in the gap
to its bulk value, remaining at any film thickness the trivial
insulator, while \btb\ at the five-SL thickness becomes a 2D TI with
the small inverted gap of 6 meV. For thicker films, the gap is zero
that indicates on the converged Dirac cone of the 3D TI phase. The
surface spectrum of latter is shown in Fig.~\ref{BTB-BTC} (d).

\section{Conclusions}

In summary, by using first-principles calculations we have examined
the topologically protected edge states and topologically trivial
Rashba-split spin-polarized dangling bond states emerged at the
different edges of the 2D topological insulator \bti\ of
one-sextuple-layer thickness. We have revealed that irrespective of
the edge plane the dangling bond states lie rather far from the 2D
band projected gap energy region, where the $\bar\Gamma$ Dirac state
resides. We also suggested the appropriate substrate for 2D \bti\
QSHI, which is parental non-centrosymmetric BiTeI compound. This
substrate guarantees the absence of both lattice mismatch and
interface charge transfer effects. Since BiTeI is the material with
the giant Rashba-type spin splitting, it produces a sizable spin
splitting in 2D band structure of QSHI film. However, despite the
Rashba splitting in the states forming the gap the latter remains
inverted, and the 2D topological phase survives in \bti/BiTeI
heterostructure as confirmed from the $\mathbb Z_2$ index
calculation. We also systematically examined the possibility of
realizing nontrivial 2D and 3D topological phases in thin films and
bulk crystals of related hypothetical compounds composed of
centrosymmetric sextuple layers of BiTeBr and BiTeCl. We have
revealed that ultra thin centrosymmetric films of \btb\ are trivial
band insulators up to five-SL thickness, and in the 3D limit \btb\
is a strong TI, while both 2D and 3D phases of \btc\ are
topologically trivial. We hope that our study will motivate further
experimental works on topologically non-trivial centrosymmetric
BiTe$X$ systems.

This work was supported by the Spanish Ministry of Economy and
Competitiveness MINECO (Project No. FIS2016-76617-P) and Saint
Petersburg State University (Grant No. 15.61.202.2015).

\end{document}